\documentclass[final,11pt,5p,twocolumn]{elsarticle}

\usepackage{graphicx}

\usepackage{color, amssymb}
\usepackage{txfonts, longtable, dcolumn, ctable, lscape, multirow, bbding, pifont, afterpage}
\usepackage{natbib}
\usepackage{microtype}
\bibpunct{(}{)}{;}{a}{}{,}

\usepackage{hyperref}

\journal{Icarus}

\begin{document}

\begin{frontmatter}

\title{Shape models of asteroids based on lightcurve observations with BlueEye600 robotic observatory}

\author[label1]{Josef \v{D}urech\corref{cor1}}
\address[label1]{Astronomical Institute, Faculty of Mathematics and Physics, Charles University, V Hole\v sovi\v ck\' ach 2, 180 00 Prague, Czech Republic}

\cortext[cor1]{Corresponding author}

\ead{durech@sirrah.troja.mff.cuni.cz}

\author[label1]{Josef Hanu\v{s}}

\author[label1]{Miroslav Bro\v{z}}

\author[label1]{Martin Lehk\'y} 

\author[label2]{Raoul Behrend}
\address[label2]{Geneva Observatory, CH-1290 Sauverny, Switzerland}

\author[label3]{Pierre Antonini}
\address[label3]{Observatoire des Hauts Patys, F-84410 B\'edoin, France}

\author[label4]{Stephane Charbonnel}
\address[label4]{Observatoire de Durtal, F-49430 Durtal, France}

\author[label5]{Roberto Crippa}
\address[label5]{Stazione Astronomica di Sozzago, I-28060 Sozzago, Italy}

\author[label4]{Pierre Dubreuil}

\author[label6]{Gino Farroni}
\address[label6]{Courbes de rotation d'ast\' ero\" ides et de com\` etes, CdR}

\author[label4]{Gilles Kober}

\author[label4]{Alain Lopez}

\author[label5]{Federico Manzini}

\author[label7]{Julian Oey}
\address[label7]{Kingsgrove, NSW, Australia}


\author[label8]{Raymond Poncy}
\address[label8]{Rue des Ecoles 2, F-34920 Le Cr\`es, France}

\author[label9]{Claudine Rinner}
\address[label9]{Ottmarsheim Observatory, 5 rue du Li\`evre, F-68490 Ottmarsheim, France}

\author[label10]{Ren\'e Roy}
\address[label10]{Observatoire de Blauvac, 293 chemin de St Guillaume, F-84570 Blauvac, France}

\begin{abstract}
We present physical models, i.e. convex shapes, directions of the rotation axis, and sidereal rotation periods, of 18 asteroids out of which 10 are new models and 8 are refined models based on much larger data sets than in previous work. The models were reconstructed by the lightcurve inversion method from archived publicly available lightcurves and our new observations with BlueEye600 robotic observatory. One of the new results is the shape model of asteroid (1663)~van~den~Bos with the rotation period of 749\,hr, which makes it the slowest rotator with known shape. We describe our strategy for target selection that aims at fast production of new models using the enormous potential of already available photometry stored in public databases. We also briefly describe the control software and scheduler of the robotic observatory and we discuss the importance of building a database of asteroid models for studying asteroid physical properties in collisional families.
\end{abstract}


\end{frontmatter}


\section{Introduction}

The increasing amount of available photometric data for asteroids has led to hundreds of asteroid shape models that have been derived from these data. A common method of asteroid shape reconstruction from disk-integrated time-resolved photometry is the lightcurve inversion of \citet{Kaa.ea:01}. The scientific motivation for reconstructing physical models of asteroids can be manifold: increasing the number of models for better statistical studies \citep{Hanus2016a}, debiasing the spin and shape distribution \citep{Mar.ea:15}, or studying the spin distribution of collisional family members \citep{Slivan2009,Hanus2013c,Kim.ea:14}, to name a few. In order to uniquely determine asteroid's sidereal rotation period, the direction of its rotation axis, and a convex shape approximation of its shape, lightcurves from different viewing and illumination geometries have to be observed. In practice it means that for a typical main-belt asteroid, we need to collect lightcurves from at least three apparitions. So to derive a new asteroid model, one has to either devote a significant amount of time to collect data from more apparitions, choose a near-Earth object that changes its geometry a lot during a single close approach, or use some archived data and combine them with new observations. For the purpose of this paper, we have chosen the last strategy and present new models of asteroids that were obtained by investing only minimum observing time and using mainly archived data.

We present our strategy of concentrating on those asteroids, for which there is a `subcritical' number of lightcurves available in the archives and adding observations from just one more apparition should lead to a unique model. A list of such candidates is published in every issue of the Minor Planet Bulletin \citep[][for example]{War.ea:16}. We used the BlueEye600 telescope that we briefly describe together with the target selection algorithm in Sect.~\ref{sec:BE600}. In Sect.~\ref{sec:results}, we present our results -- new or updated shape models of 18 asteroids.
lightcurve
\label{sec:BE600}

New lightcurves were obtained with the robotic observatory BlueEye600. 
We have developed an algorithm to assign priorities to individual asteroids, with the aim to maximize the number of new asteroid models that could be derived by adding new lightcurves to archived data. Therefore, we focused on asteroids for which older lightcurves existed and scheduled the observations according to priorities of observable asteroids and observing conditions. 


\subsection{Instrument description}

The observatory is located in Ond\v rejov, Czech Republic ($\phi = 49^\circ54'34''$, $\lambda = 14^\circ46'48''$, $h = 515\,{\rm m}$).
The telescope itself is a Ritchey--Chr\'etien system (Officina Stellare),
with the primary mirror diameter $d = 600\,{\rm mm}$
and the effective focal length $f = 3\,000\,{\rm mm}$,
equipped with a 3-lens optical corrector.
It produces a diffraction-limited images in a~large field of view (50\,mm),
after a~proper alignment; in practice the telescope is seeing limited.
Seeing conditions at the site are not exceptional, the average value is about 2--3\,arcsec.
The secondary mirror also contains a 2nd (raw) focuser.

Instruments are located in the secondary focus, in particular
focuser,
derotator (Rotofocuser)
off-axis guider with a 2nd camera
filter wheel (FLI)
and the main camera (Mii) using a~E2V~42--40 CCD chip,
with more than 90\,\% quantum efficiency in VRI bands
and the resulting FOV $0.52^\circ$.

The alt-azimuth mount (developed by Projectsoft)
is of very rigid construction, equipped with torque motors
and allows for fast motions with the angular velocity up to $45^\circ/{\rm s}$
and angular acceleration up to $45^\circ/{\rm s}^2$.

\begin{figure}
\centering
\includegraphics[width=\columnwidth]{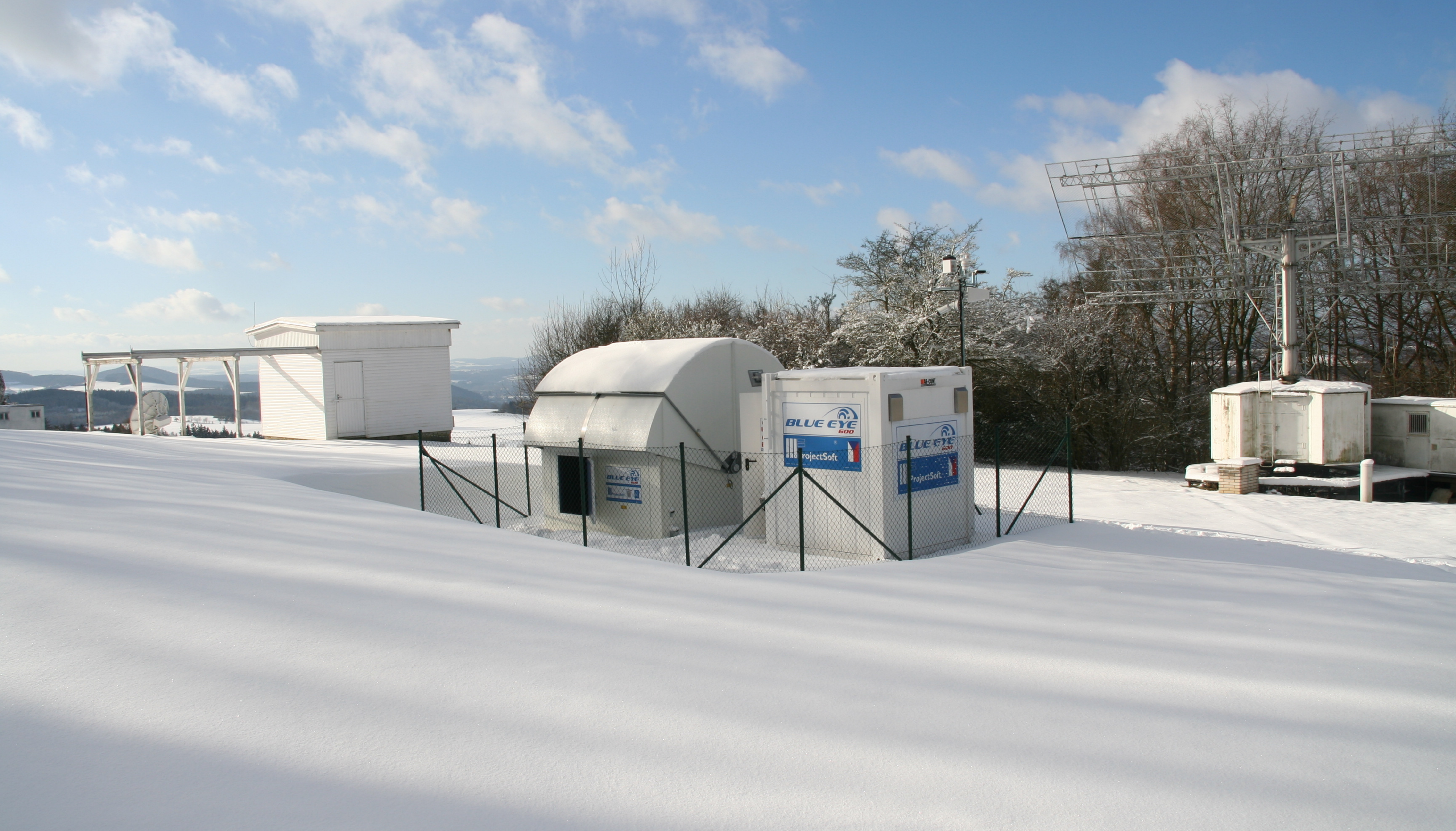}
\caption{Projectsoft BlueEye600 robotic observatory located in Ond\v rejov, Czech Republic.
The alt-azimuth telescope is of Ritchey--Chr\'etien type with the primary diameter $d = 600\,{\rm mm}$.}
\label{blueeye_img_9374}
\end{figure}


\subsection{High-level control software}

\begin{figure}
\centering
\includegraphics[width=\columnwidth]{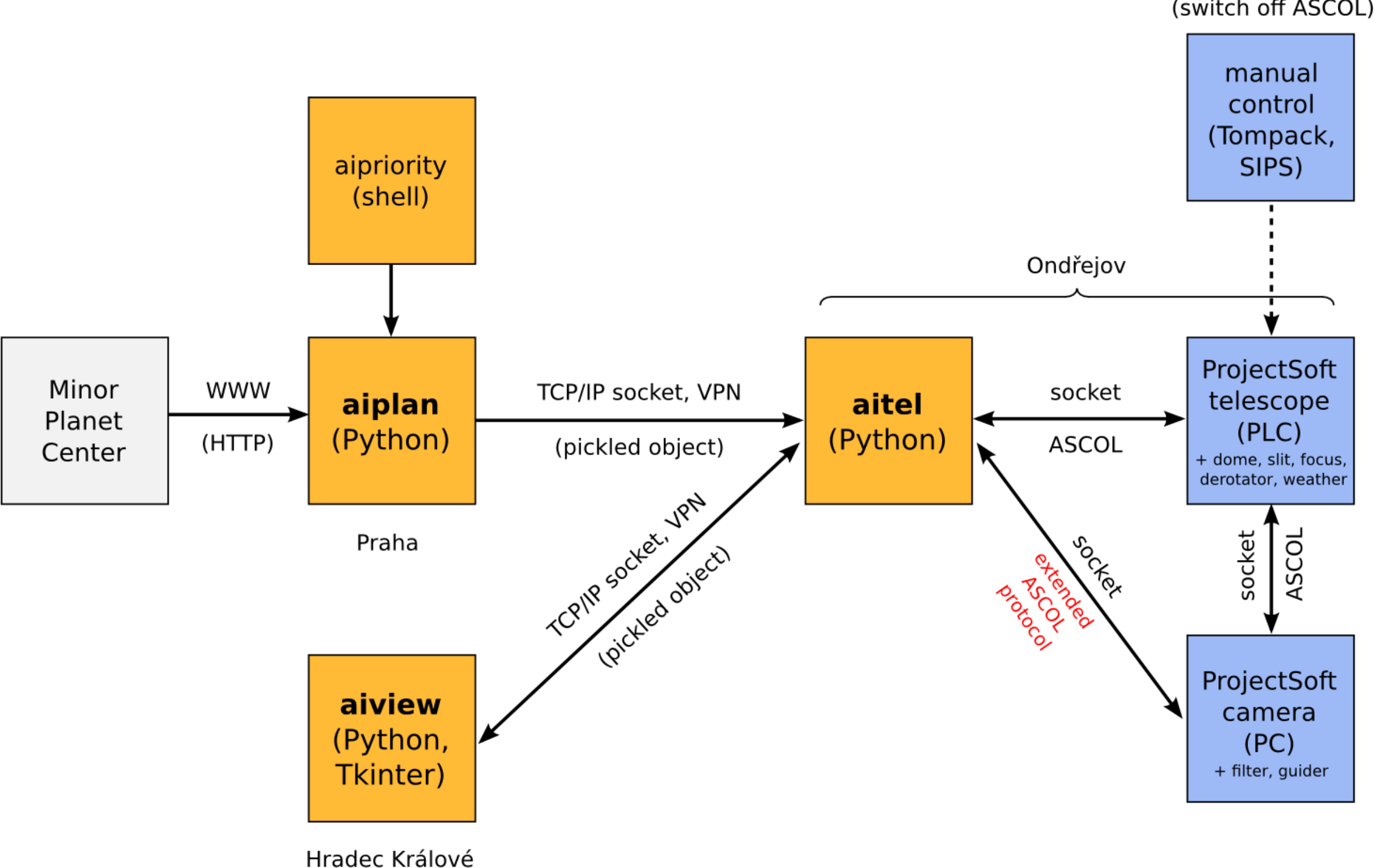}
\caption{A concept scheme of the high-level software for autonomous control
of the observatory. The basic components are:
Aitel (executive part controlling the observatory),
Aiplan (automated planner of observations),
Aiview (graphical user interface for manual planning of observations).
The ephemerides are downloaded from the Minor Planet Center.
On the other hand, there is lower-level control software
(Projectsoft telescope and Projectsoft camera),
which receives commands over TCP/IP according to ASCOL and MACOL protocols.
It also allows for a manual control or debugging.}
\label{aitel_scheme3}
\end{figure}

The observatory can be controlled using a high-level software
with the following basic architecture or parts:
Aitel, Aiplan, Aiview, Projectsoft telescope and~Projectsoft camera
(see the simplified scheme in Figure~\ref{aitel_scheme3}).
All communication between these software components is based on TCP/IP sockets.
One can either send commands and their parameters to the lower-level software,
specified according to ASCOL, MACOL and other protocols, or serialised objects
(most importantly `cubes', i.e. descriptions of observational blocks).
The communication is transparent over the internet, or a secure VPN network, respectively.

\begin{figure}
\centering
\includegraphics[width=\columnwidth]{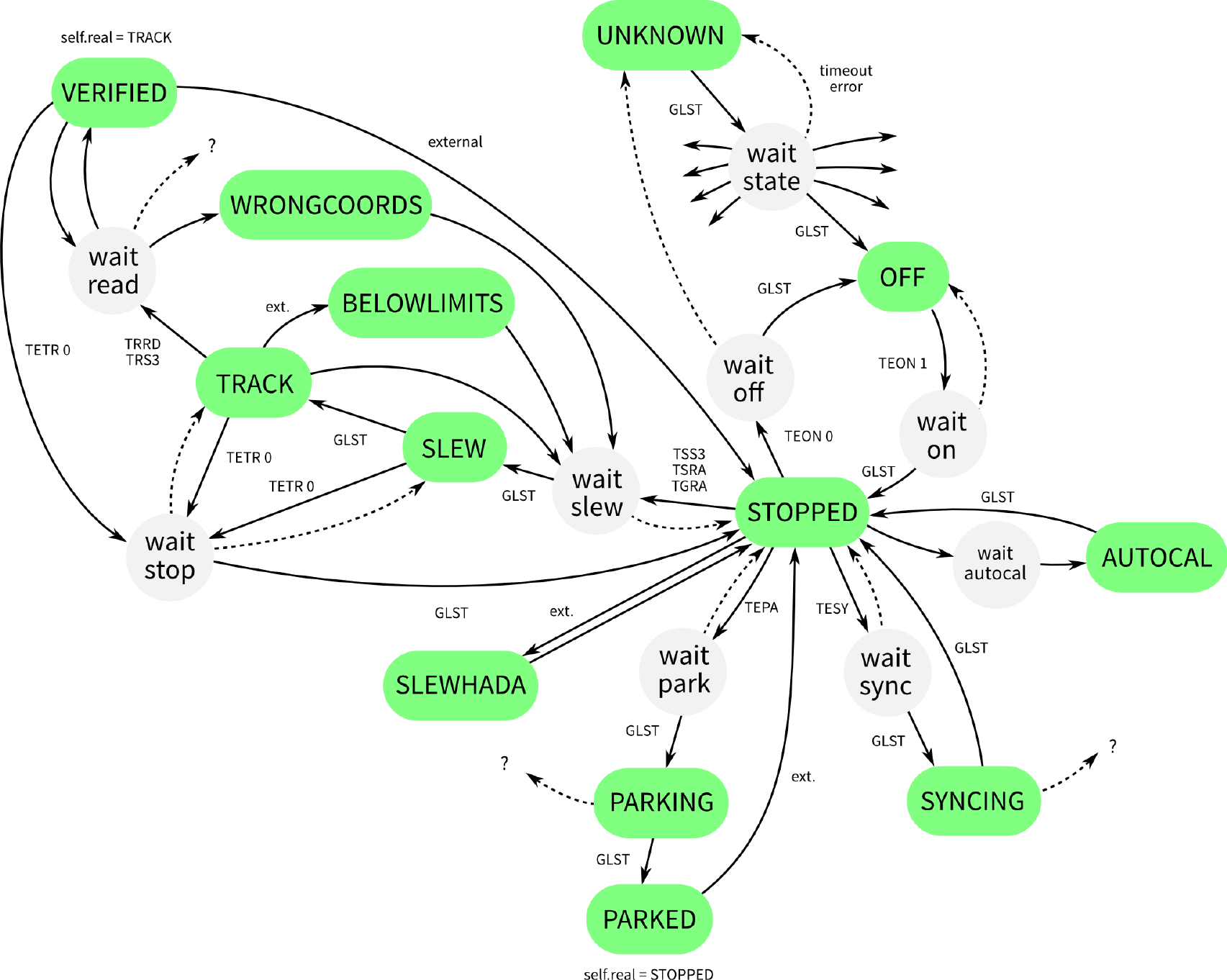}
\caption{A block scheme of telescope functions which
serves as a guide for the implementation of Tel (i.e. telescope) software class.
All blocks correspond to 22 different states in which the object can be.
Green blocks are the states to which the object can move on its own,
after receiving a reply on GLobal STate (GLST) command.
Gray blocks are transition-states in which the object is just after
sending an executive command, according to the ASCOL protocol (see the respective small labels).
The arrows correspond to possible transitions,
dashed arrows indicate changes after a timeout,
or an errorneous reply.
The object appears as separate, as being independent of others,
but all objects are interconnected in a different part of the code.
called the sequencer.}
\label{tel}
\end{figure}

The most important component is Aitel, the executive part
running in real time and sending commands to the observatory.
It is implemented in Python language as an object-oriented
and fully asynchronous code. Functional schemes exist for all types of devices,
which appear as classes in the code (see the example of Tel class in Figure~\ref{tel}).
For every individual device there is an object, the instance of the corresponding class.
The object-oriented language allows for an easy modifications and future development
of other scientific applications.

The list of supported classes corresponding to types of hardware devices includes:
Autofocus, Camera, CameraVoltage, Converter, Cooling,
Derotator, Dome, Filterwheel, Flap, Focus, Guider, Heating, Meteo, Meteodata,
Slit, Tel, Voltage, Weather.
Apart from them, we also have classes serving for configuration and TCP/IP communications:
AscolClient, Config, CubeClient, CubeServer, Macol\-Client.
There might be more objects of the same class. 
In particular, we use three objects ascol\_client, i.e. instances of AscolClient class.
The source code of the given class is devoted exclusively to a single device,
because mutual dependencies are accounted for elsewhere --- in the sequencer code.
This strict distinction makes the code very clean.

At the same time, the code is fully asynchronous, so there are
no waiting cycles that would delay reaction of the program on changes
of object states. The TCP/IP communications can be generally fragmented
and we use the select module for sending and receiving packets,
which minimizes the CPU load.

\subsection{Planning of observations}

In general, observable asteroids that are within the reach of the telescope are sorted according to their priority, that is assigned according to the preferences. The highest priority is given to those asteroids for which there are old lightcurves from more apparitions and still no unique model. In such cases, we expect to derive a unique model by enlarging the data set by just one or a few lightcurves from a new apparition. We also gave high priority to those asteroids for which a preliminary model existed but was based only on low-quality sparse photometry \citep[][for example]{Dur.ea:16}. The confirmation of models based on sparse data from surveys is important because sparse data will dominate the field of lightcurve inversion just by sheer amount of data. By comparing models derived from sparse data with those from dense lightcurves, we will understand limitations of models based on sparse photometry.

The weight of the observation is a number $w \in [0; 10]$ from the given interval,
which expresses our priority for the observation of the given asteroid.
The respective computation is performed in Aiplan
and the value is stored in the `cube'.
The user can check the cube (and its weight) and possibly modify it with Aiview.
Finally, Aitel sorts all the cubes according to the weight,
in descending order, and selects the first one which is observable
at the given time instant.

We use the following equation to compute the weight:
\[w = w_{\rm d}'\, f_{\rm airmass}\, f_{\rm duration}\, f_{\rm moonalt}\, f_{\rm elong}\, f_{\rm phase}\, f_{\rm eclipse}\, f_{\rm sn}\, f_{\rm ucac}\,,\]
where $w_{\rm d}'$ denotes the weight according to our database of existing observations,
which is independent of observational conditions,
multiplied by factors $f \in [0; 1]$ dependent on observations.
Namely
$f_{\rm airmass}$~denotes the factor corresponding to air mass (or height above the horizon),
$f_{\rm duration}$~duration of the observation in comparison with the rotation period,
$f_{\rm moonalt}$~height of the Moon,
$f_{\rm elong}$~elongation of the Moon,
$f_{\rm phase}$~phase of the Moon,
$f_{\rm eclipse}$~an eclipse by the Moon,
$f_{\rm sn}$~signal-to-noise ratio
and~$f_{\rm ucac}$~encounters between the asteroid and stars according to UCAC4 catalogue \citep{Zac.ea:13}.


\subsubsection{Weigths according to the database}

The weights according to our light curve database and eventually Brian Warner's database of periods (Warner et al. 2009)
are assigned as follows:
\begin{itemize}
\setlength\itemsep{0pt}
    \item [10 --] there is a non-physical shape model, the asteroid is on a `hot list'
    \item [9 --]  `hot list' with a slightly lower weight
    \item [8 --]  known period, more poles, $>4$~apparitions, `hot list'
    \item [7 --]  known period, more poles, 2~to 4~apparitions
    \item [6 --]  known period, more poles, single apparition, or unknown period, $>4$~apparitions
    \item [5 --]  known period, only sparse data, the number of observations $\ge 100$, or unknown period, $4$~apparitions
    \item [4 --]  unknown period, 2~or 3~apparitions
    \item [3 --]  known period, only sparse data, the number of observations 70 to~99, or unknown period, 2~or 3~apparitions
    \item [2 --]  unknown period, only sparse data, the number of observations $\ge 100$
    \item [1 --]  unknown period, only sparse data, the number of observations 70 to~99
    \item [0 --]  there is already a model in DAMIT database, unpublished model
\end{itemize}

The values of $w_{\rm d}$ can be further increased or decreased by $w_i$
in the following specific cases:
$w_1 = +3$: asteroid is a member of a major asteroid family \citep{Nes:12, Nes:15};
$w_2 = +2$: an ellipsoidal model exists;
$w_3 = -3$: the asteroid is binary.
The resulting weigth is then:
$$w_{\rm d}' = {\rm min}\left(w_{\rm d} + \sum w_i, 10\right)\,.$$


\subsubsection{Factors depending on observational conditions}

Factors~$f$ are computed from the asteroid ephemeris and Moon ephemeris
downloaded from the Minor Planet Center (MPC). We use a whole arc
for this purpose, not only extremal values.
The factor corresponding to air mass, or the height above the horizon, respectively,
is defined as:
$$f_{\rm airmass} = {\int_t H(h)\sin h\,{\rm d}t \over \int_t{\rm d}t }\,,\label{f_airmass}$$
where $t$ denotes time (Julian date),
$h(t)$~height of the asteroid
and~$H(x)$~Heaviside step function.
The factor would be
$f_{\rm airmass} = 1$, if there would be $h = 90^\circ$ during the whole observation (a kind of miracle),
and~$f_{\rm airmass} = 0$ for $h \le 0^\circ$.

The factor of duration is:
$$f_{\rm duration} = {\rm min}\left[{\int_t H(h-h_{\rm min})\,{\rm d}t\over P/4}, 1\right]\,,$$
where~$h_{\rm min}$ denotes the minimum height of the target (a technological limit or the local horizon)
and~$P$~the rotation period, if known; otherwise $f_{\rm duration} = 1$.

\newcommand{\mesic}{{\rm C}}
The factor of Moon height is:
$$f_{\rm moonalt} = 1 - {\int_t H(h_\mesic)\sin h_\mesic\, k_{\rm full} f_\mesic\, H(h-h_{\rm min})\,{\rm d}t \over \int_t H(h-h_{\rm min})\,{\rm d}t }\,,$$
where~$h_\mesic$ denotes the angular height of the Moon
and~$f_\mesic$ its phase.
It is an equivalent of Eq.~\ref{f_airmass}, but only for a limited time span, when the asteroid of interest is above the horizon.
We also see, that a new Moon will not affect us ($f_\mesic = 0$).
The coefficient $k_{\rm full} \simeq 0{,}9$ is not unity,
because we shall observe even during a full Moon in the zenith.

The factor of Moon elongation is:
$$f_{\rm elong} = {\int_t [ {1\over 2}(1-\cos e_\mesic) H(h_\mesic) f_\mesic + H(-h_\mesic) ] H(h-h_{\rm min}) \,{\rm d}t \over \int_t H(h-h_{\rm min})\,{\rm d}t}\,,$$
where~$e_\mesic$ is the elongation of the Moon and the asteroid.
These combinations of the Heaviside functions guarantee that the factor is independent of $e_\mesic$,
if it is not above the horizon.

The factor of Moon phase is:
$$f_{\rm phase} = {\int_t [(1-k_{\rm full}f_\mesic) H(h_\mesic) + H(-h_\mesic)] H(h-h_{\rm min})\,{\rm d}t \over \int_t H(h-h_{\rm min})\,{\rm d}t}\,,$$
where $f_\mesic$ denotes the phase of the Moon,
i.e.~1 for the full Moon and 0~for the new one.

The factor of an Moon eclipse, even thought quite improbable, is:
$$f_{\rm eclipse} = {\int_t H(e_\mesic-\alpha_\mesic) H(h-h_{\rm min})\,{\rm d}t \over \int_t H(h-h_{\rm min})\,{\rm d}t}\,,$$
where $\alpha_\mesic$ is the angular radius of the Moon (constant here).
This approach can be also used when we want to elliminate possible reflections
due to the Moon then we choose larger $\alpha_\mesic \simeq 1^\circ$).

The factor corresponding to the ratio ${{\rm S}\over{\rm N}}$,
i.e. the signal from asteroid over the noise from all possible sources,
computed for the given maximal exposure time~$t_{\rm exp} \simeq 180\,{\rm s}$, is:
$$f_{\rm sn} = H_3\left\{ \max\left[ \min\left({{{\rm S}\over{\rm N}} - {{\rm S}\over{\rm N}}\big|_{\rm min} \over {{\rm S}\over{\rm N}}\big|_{\rm max} - {{\rm S}\over{\rm N}}\big|_{\rm min} }, 1\right), 0\right] \right\}\,,$$
where
$H_3(x) \equiv 3x^2 - 2x^3$ is the Hermite polynomial of the third order;
we choose some limiting values
${{\rm S}\over{\rm N}}\big|_{\rm min} \simeq 5$,
${{\rm S}\over{\rm N}}\big|_{\rm max} \simeq 100$.

A~`cube' for observation is generated only if the resulting $w>0$, of course.

\subsection{Observations and reductions}

Photometric observations were mostly performed in the standard Kron-Cousins Rc filter. Exposure times were between 60 and 180\,s, depending on the brightness of the target. The limiting magnitude is about 18\,mag for the longest exposures. We rather prefer to observe one or a few targets per night with a high cadence to obtain a high total signal (and S/N) and to significantly constrain the shape modelling. Our new light curve observations (together with older ones) are summarised in Table~\ref{tab:references}. In 2016, we observed 45 different objects for about 227 hours, and 26 additional objects in the same field of view for 105 hours. In 2015, it was 11 objects for 197 hours. Only a subset of these data were used in this work, thought, because for many asteroids observed by BlueEye600 we still do not have enough data to derive a unique shape model.

Basic reduction procedures are not automated. We use a standard series of
bias, dark-frame and flat-field corrections. Optionally, we can apply a procedure to suppress fringing if present \citep{2013Msngr.152...14S}. We then perform an aperture photometry with C-Munipack\footnote{\url{http://c-munipack.sourceforge.net}} vers. 1.1.26 software by Motl et al., essentially based on the same principles as Daophot \citep{2011ascl.soft04011S}. The matching algorithm is robust \citep{1998stel.conf...30H}; it was only adjusted to better work with moving objects, which have to be manually pin-pointed on two images. The uncertainties of the measurements are calculated as poissonian photon noise, with contributions from the sky, dark, readout, and discretisation. For the subsequent analysis, only relative lightcurves are needed and we thus do not perform an absolute calibration.

\section{New and updated asteroid shape models}
\label{sec:results}

When combining our new observations with archived data, we followed the same approach as \cite{Hanus2016a}. We used lightcurves archived in the Asteroid Photometric Catalogue \citep{Piironen2001} and the Asteroid Lightcurve Photometry Database\footnote{\url{http://alcdef.org}} \citep{Warner2011d}. We also used sparse photometry from the US Naval Observatory in Flagstaff, Catalina Sky Survey Observatory (downloaded from the AstDyS), and the Lowell Photometric Database \citep{Osz.ea:11, Bow.ea:14}. The number of individual lightcurves and sparse data points for each asteroid are listed in Table~\ref{tab:models}, the sources of dense lightcurves are listed in Table~\ref{tab:references}. For each asteroid, we searched for the best sidereal rotation period and corresponding solution for the pole direction and shape model. The initial search interval was set according to the synodic period listed in the LCDB database of \cite{War.ea:09}.

The list of asteroid models is given in Table~\ref{tab:models}. For each asteroid, we give the direction of its spin axis in ecliptic coordinates, the sidereal rotation period, and the number of lighcurves, apparitions, and sparse data points in the data set we used. Because asteroids orbit close to the plane of the ecliptic, the geometry is restricted to this plane and there are usually two pole solutions that provide the same fit to the data \citep{Kaa.Lam:06}. The shape models, figures showing the fit to lightcurves, and photometric data used for the model reconstruction are available through the Database of Asteroid Models from Inversion Techniques \citep[DAMIT\footnote{\url{http://astro.troja.mff.cuni.cz/projects/asteroids3D}},][]{Dur.ea:10}. The new lightcurves observed with the BlueEye600 telescope were also uploaded to the Asteroid Lightcurve Photometry Database$^1$ \citep{Warner2011d}.

In total, we reconstructed 18 asteroid models, out of which 10 are new models. The remaining 8 are updated models of those based only on sparse photometry from the Lowell Photometric Database and published by \cite{Dur.ea:16}. In general, the updated spin solutions agree within the expected errors with those based on only sparse data, which further enhances reliability of models reconstructed from sparse photometry. The mean difference of poles between new and old models of the same asteroid is $\sim 16^\circ$ of arc. The largest discrepancies are around $25^\circ$ with only one exception for the second pole solution for (1320)~Impala where the spin axis directions for the two models differ by about $35^\circ$.

\begin{figure}
\centering
\includegraphics[width=\columnwidth]{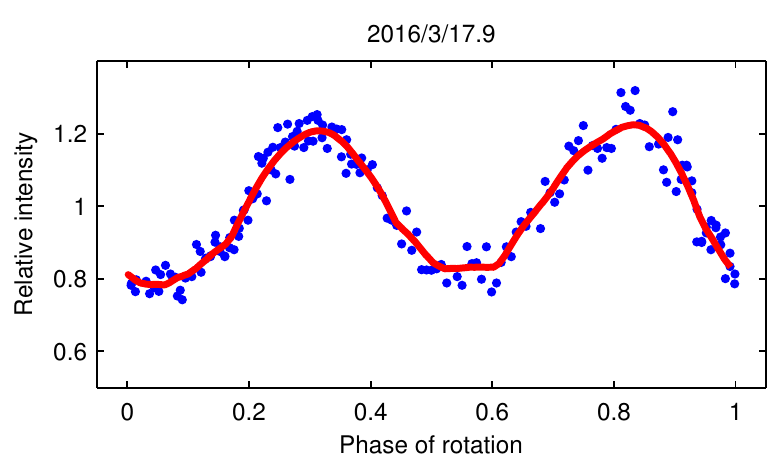}\\
\medskip
\includegraphics[width=\columnwidth]{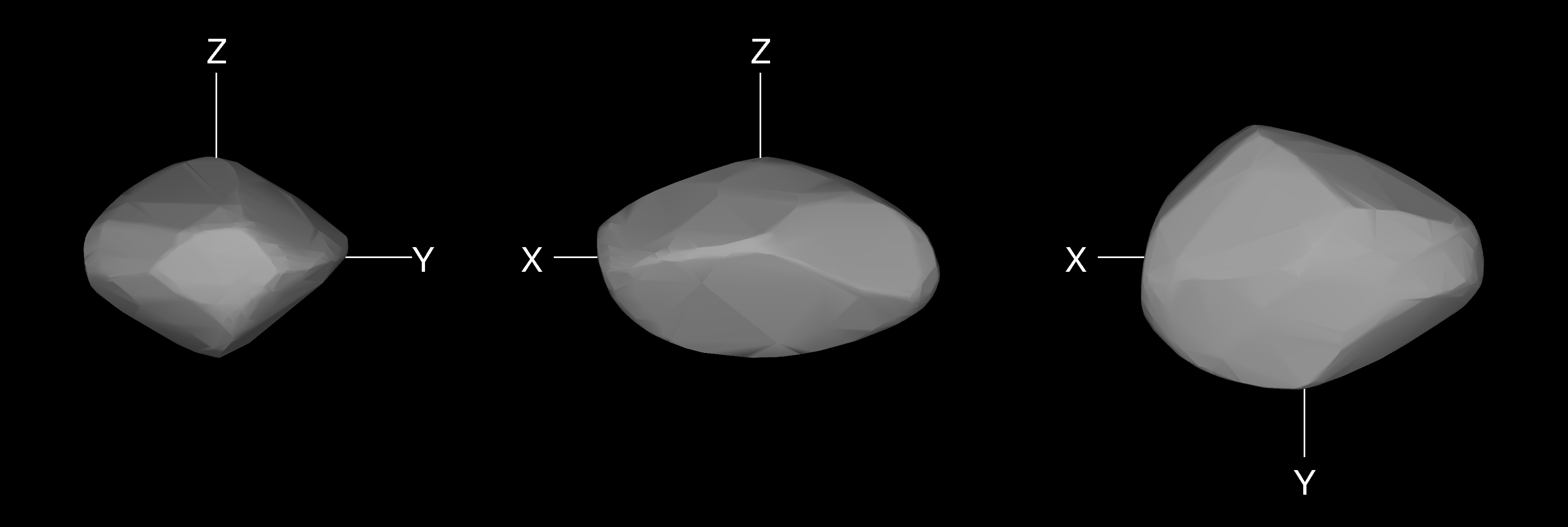}
\caption{Lightcurve of asteroid (955)~Alstede observed by BlueEye600 and the synthetic lightcurve (top) produced by the shape model (bottom) that was reconstructed from this lightcurve and other 258 sparse photometric measurements.}
\label{fig:955_lcfit}
\end{figure}

As an example, we show in Fig.~\ref{fig:955_lcfit} an updated model of asteroid (955)~Alstede, that was uniquely (up to the ambiguity in the pole longitude) reconstructed from one lightcurve and sparse data.

There are several members of asteroid families in our small sample. Most importantly, (163) Erigone is the largest remnant of the corresponding C/X-type family \citep[designated FIN 406, according to Family Identifier Number defined by][]{Nes.ea:15}, in the inner belt.
A very interesting case is (918)~Itha, the largest fragment of the S-type Itha family (FIN 633), located in the so-called `pristine zone' of the main belt (2.832 to 2.956\,au). This family has an extremely shallow size-frequency
distribution (cumulative slope $q = 1.3 \pm 0.3$) which may indicate this is actually a remnant of a very old disruption originated during the Late Heavy Bombardment \citep{Bro.ea:13}.
There are also three members of the Eos family (FIN 606) in our sample: (608)~Adolfine, (775)~Lumiere, and (1095)~Tulipa;
one member of Maria family (FIN 506): (616)~Elly; and one from Vesta (FIN 401): (2511)~Patterson.

\begin{figure}
\centering
\includegraphics[width=\columnwidth]{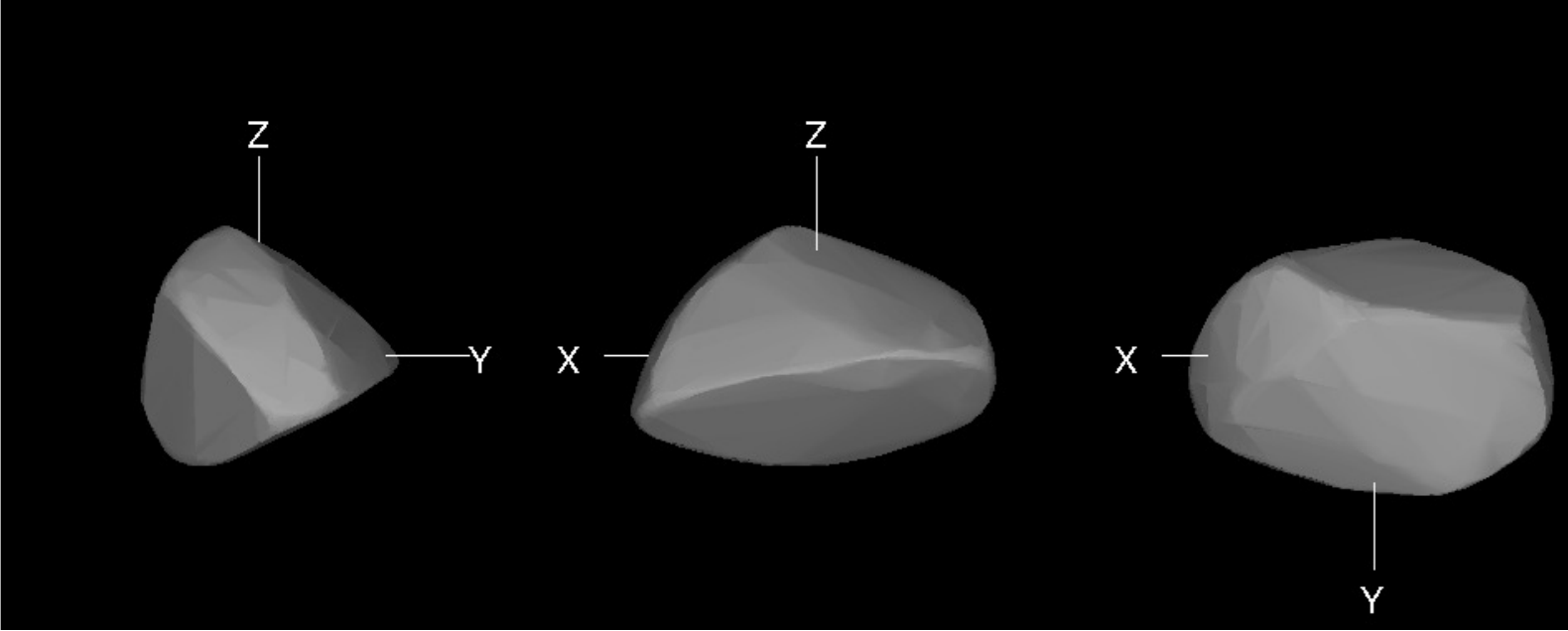}
\caption{Shape model of asteroid (1663) van den Bos for the pole direction $(248^\circ, 62^\circ)$ shown from two equatorial views (left and center) and pole-on (right).}
\label{fig:1663_model}
\end{figure}

\paragraph{The slowest rotator with known shape}
One of the new models -- (1663)~van den Bos -- is particularly interesting because it has a very long rotation period of 748.7\,hr, which is close to the synodic period of $740 \pm 10$\,hr determined by \cite{Stephens2011c}. Although we cannot rule out some excitation of its rotation state, it cannot be very high because we obtained a model with the same period and similar pole directions from inversion of an independent set of calibrated sparse data from the Lowell Photometric Database. The shape model for one of the possible poles is shown in Fig.~\ref{fig:1663_model}.

This asteroid is an S-type, with the geometrical albedo $p_V = 0.171\pm 0.018$
and diameter $D = (11.70\pm0.05)\,{\rm km}$ \citep{Masiero2011},
The time scale of YORP-effect-driven evolution (or doubling time) can be estimated as:
\[\tau_{\rm YORP} = \tau_0 c_{\rm YORP}^{-1} \left({a\over a_0}\right)^{\!2} \left({D\over D_0}\right)^{\!2} \left({\rho\over\rho_0}\right) \doteq 990\,{\rm Myr}\,,\]
where
$\tau_0 = 11.9\,{\rm Myr}$,
$a_0 = 2.5\,{\rm au}$,
$D_0 = 2\,{\rm km}$,
$\rho_0 = 2\,500\,{\rm kg}\,{\rm m}^{-3}$
are the reference values from \cite{Capek2004}.
Note that for decreasing the spin rate from an average value
of $D = 10\,{\rm km}$ asteroids, $\bar\omega \simeq 5\,{\rm rev}\,{\rm day}^{-1}$,
down to the current value, one would need about 7~doubling times.

At the same time, we can estimate a collisional reorientation time scale as:
\[\tau_{\rm reor} = B\left({\omega\over\omega_0}\right)^{\beta_1} \left({D\over D_0}\right)^{\beta_2} \doteq 140\,{\rm Myr}\,,\]
where we used
$B = 84.5\,{\rm kyr}$,
$\omega_0 = 2\pi/P_0$,
$P_0 = 5\,{\rm h}$,
$\beta_1 = 5/6$, and
$\beta_2 = 4/3$
according to \cite{Farinella1998}.
It means that the asteroid spin is currently rather driven by collisions than YORP.
However, if the rotation was significantly faster in the past,
the collisional time scale was most likely longer than $\tau_{\rm YORP}$.

The time scale for damping of non-principal-axis rotation is of the order of \citep{Hestroffer2006}:
\[\tau_{\rm damp} = {\mu Q\over\rho K_1^2 (D/2)^2\omega^3} \simeq 10^6\,{\rm Myr}\,,\]
where we assumed
$\rho = 2\,500\,{\rm kg}\,{\rm m}^{-3}$,
$K_1^2 \simeq 0.1$, and
the value $\mu Q = 5\times10^{12}\,{\rm Pa}$ \citep{Harris1994}.
In the current state, the damping thus seems negligible
and from this point of view one would expect an excited state.

\section{Conclusions}

The new models we have obtained will further increase the number of asteroids for which a simple physical model exists. By building a database of such models, we can study the distribution of spins and shapes across the population. We will re-observe those targets, for which the current data sets are still not enough to build a unique model.  The data obtained in the course of this project (together with other data) will be also used as independent observational constraints for dynamical models of asteroid families and their longer-term evolution, driven mostly by the Yarkovsky semimajor axis drift, chaotic diffusion in mean-motion resonances, eccentricity and inclination drift in secular resonances \citep{Broz2013a}, random collisional reorientations, the YORP effect causing systematic changes of spin rates and spin poles \citep{Hanus2013c}, or even spin-orbital resonances \citep{Vokrouhlicky2003}.

Luckily, the current number of shape models in DAMIT database seems
sufficiently high, in order to focus on individual families. For example,
recent census of the Eos family shows there are at least 43 core and 27 halo
asteroids (including background) with known spin orientations (Hanu\v{s} et al., in prep.). There are three more Eos family members in our list of models. Together with distributions inferred from Lowell photometric data, in particular the absolute
values of (approximate) pole latitudes $|\beta|$ for 69\,053 asteroids \citep{Cibulkova2016},
they may represent very stringent constraints for dynamics of small ($D \simeq 10^0\,{\rm km}$) asteroids.

\section*{Acknowledgements}
The work of MB and J\v{D} was supported by the Grant Agency of the Czech Republic (grant no. 15-04816S). The computations have been done on the computational cluster Tiger at the Astronomical Institute of Charles University in Prague (\url{http://sirrah.troja.mff.cuni.cz/tiger}).
The development of the observatory was supported by the Technology Agency of the Czech Republic (TA \v CR),
project no.~TA 03011171, 'Development of Technologies for the Fast Robotic Observatories
and Laser Communication Systems'.

\onecolumn
\setlength{\tabcolsep}{5pt}
\begin{table}
\caption{\label{tab:models}Rotational states and light curve summary of asteroids, for which we derived their shape models. The table gives ecliptic coordinates $\lambda_1$, $\beta_1$, $\lambda_2$ and $\beta_2$ of the two best-fitting pole solutions, sidereal rotational period $P$, the number of dense light curves $N_{\mathrm{lc}}$ spanning $N_{\mathrm{app}}$ apparitions, the number of sparse-in-time measurements $N_{\mathrm{sp}}$, and the reference to the model.}
\begin{tabular}{r@{\,\,\,}l rrrr D{.}{.}{6} rrrc}
\hline
\multicolumn{2}{c} {Asteroid} & \multicolumn{1}{c} {$\lambda_1$} & \multicolumn{1}{c} {$\beta_1$} & \multicolumn{1}{c} {$\lambda_2$} & \multicolumn{1}{c} {$\beta_2$} & \multicolumn{1}{c} {$P$} & $N_{\mathrm{lc}}$ & $N_{\mathrm{app}}$  & $N_{\mathrm{sp}}$ & \multicolumn{1}{c}{Reference} \\
\multicolumn{2}{l} { } & [deg] & [deg] & [deg] & [deg] & \multicolumn{1}{c} {[hours]} &  &  &  & \multicolumn{1}{c}{}\\
\hline\hline
114    &  Kassandra  &  196  &  $-$55  &  4    &  $-$58  &  10.74358   &  24   &  8  &  322  &  This work           \\
163    &  Erigone    &  191  &  $-$75  &  358  &  $-$73  &  16.1403    &  5    &  2  &  331  &  This work           \\
       &             &  276  &  $-$69  &       &         &  16.1402    &       &     &  483  &  \citet{Dur.ea:16}  \\
176    &  Iduna      &  156  &  78     &  85   &  29     &  11.28783   &   9   &  3  &  360  &  This work           \\
       &             &  219  &  68     &  83   &  24     &  11.28785   &       &     &  491  &  \citet{Dur.ea:16}  \\
582    &  Olympia    &  135  &  3      &       &         &  36.3635    &  104  &  6  &  443  &  This work           \\
608    &  Adolfine   &  342  &  43     &  171  &  30     &  8.34489    &  10   &  3  &  224  &  This work           \\
616    &  Elly       &  262  &  54     &  353  &  85     &  5.29770    &  19   &  3  &  284  &  This work           \\
       &             &  250  &  44     &  60   &  62     &  5.29770    &       &     &  368  &  \citet{Dur.ea:16}  \\
775    &  Lumiere    &  89   &  54     &  248  &  47     &  6.10300    &  14   &  6  &  250  &  This work           \\
822    &  Lalage     &  343  &  $-$74  &  133  &  $-$75  &  3.346503   &  13   &  3  &  445  &  This work           \\
918    &  Itha       &  72   &  $-$54  &       &         &  3.473808   &   9   &  2  &  227  &  This work           \\
       &             &  59   &  $-$59  &  249  &  $-$72  &  3.47381    &       &     &  350  &  \citet{Dur.ea:16}  \\
955    &  Alstede    &  64   &  57     &  246  &  23     &  5.18734    &  1    &  1  &  258  &  This work           \\
       &             &  54   &  38     &  240  &  13     &  5.18735    &       &     &  401  &  \citet{Dur.ea:16}  \\
1095   &  Tulipa     &  142  &  40     &  349  &  56     &  2.787153   &  11   &  3  &  356  &  This work           \\
1219   &  Britta     &  61   &  $-$62  &  223  &  $-$68  &  5.57557    &  21   &  3  &  231  &  This work           \\
       &             &  72   &  $-$66  &  241  &  $-$66  &  5.57556    &       &     &  387  &  \citet{Dur.ea:16}  \\
1251   &  Hedera     &  271  &  $-$53  &  115  &  $-$62  &  19.9020    &  10   &  2  &  289  &  This work           \\
       &             &  266  &  $-$62  &  124  &  $-$70  &  19.9021    &       &     &  414  &  \citet{Dur.ea:16}  \\
1320   &  Impala     &  186  &  $-$43  &  126  &  $-$70  &  6.17081    &  9    &  2  &  213  &  This work           \\
       &             &  151  &  $-$57  &  254  &  $-$70  &  6.17081    &       &     &  353  &  \citet{Dur.ea:16}  \\
1380   &  Volodia    &  82   &  54     &  261  &  29     &  6.19570    &  1    &  1  &  197  &  This work           \\
1663   & van den Bos &  248  &  62     &  43   &  51     &  748.72     &  37   &  2  &  241  &  This work           \\
2511   &  Patterson  &  194  &  50     &  10   &  31     &  4.14065    &  1    &  1  &  164  &  This work           \\
34817  & Shiominemoto & 201  &  36     &       &         &  6.37750    &  12   &  5  &  85   &  This work           \\ \hline
\end{tabular}
\end{table}
\twocolumn

\onecolumn

\begin{longtable}{r@{\,\,\,}l lll}
\caption{\label{tab:references}Light curve observations used for the shape model determinations. Our observations with the BE600 telescope were obtained by Martin Lehk\'y.}\\
\hline
 \multicolumn{2}{c} {Asteroid} & Date & $N_{\mathrm{LC}}$ & Observer/Telescope \\ \hline\hline

\endfirsthead
\caption{continued.}\\

\hline
 \multicolumn{2}{c} {Asteroid} & Date & $N_{\mathrm{LC}}$ & Observer/Telescope \\ \hline\hline
\endhead
\hline
\endfoot
  114 &       Kassandra & 1979 03 --  1980 07 &  2 & \citet{Harris1989a} \\
      &                 & 1981 09 --  1981 09 &  2 & \citet{Harris1992a} \\
      &                 & 1988 04 --  1988 04 &  3 & \citet{Hutton1988} \\
      &                 & 1993 05 --  1993 05 &  2 & \citet{Piironen1998} \\
      &                 & 2006-6-24.0         &  1 & Roberto Crippa, Federico Manzini \\
      &                 & 2006-6-27.0         &  1 & Roberto Crippa, Federico Manzini \\
      &                 & 2006-6-30.0         &  1 & Roberto Crippa, Federico Manzini \\
      &                 & 2006-7-11.0         &  1 & Roberto Crippa, Federico Manzini \\
      &                 & 2006-7-21.0         &  1 & Roberto Crippa, Federico Manzini \\
      &                 & 2006-7-21.9         &  1 & Roberto Crippa, Federico Manzini \\
      &                 & 2006-8-06.9          &  1 & Roberto Crippa, Federico Manzini \\
      &                 & 2006-8-08.9          &  1 & Roberto Crippa, Federico Manzini \\
      &                 & 2011-8-29.9         &  1 & Ren\'e Roy \\
      &                 & 2011-8-31.0         &  1 & Ren\'e Roy \\
      &                 & 2011-9-17.0         &  1 & Ren\'e Roy \\
      &                 & 2011-9-19.0         &  1 & Ren\'e Roy \\
      &                 & 2016-08-02.1        &  1 & Martin Lehk\'y \\
      &                 & 2016-08-14.0        &  1 & Martin Lehk\'y \\
      &                 & 2016-08-18.0        &  1 & Martin Lehk\'y \\
  163 &         Erigone & 1980 12 --  1981 01 &  3 & \citet{Harris1989a} \\
      &                 & 2016-11-20.8        &  1 & Martin Lehk\'y \\
      &                 & 2016-12-29.8        &  1 & Martin Lehk\'y \\
  176 &           Iduna & 2007 09 --  2007 10 &  3 & \citet{Warner2008d}\\
      &                 & 2015-01-07.7        &  1 & Julian Oey \\
      &                 & 2015-01-14.7        &  1 & Julian Oey \\
      &                 & 2015-01-17.7        &  1 & Julian Oey \\
      &                 & 2015-01-23.5        &  1 & Julian Oey \\
      &                 & 2015-01-29.6        &  1 & Julian Oey \\
      &                 & 2016-04-29.9        &  1 & Martin Lehk\'y \\
  582 &         Olympia & 1986 01 --  1989 12 & 10 & \citet{Schober1993} \\
      &                 & 2006-11-13.6        &  1 & Julian Oey \\
      &                 & 2006-11-21.6        &  1 & Julian Oey \\
      &                 & 2013 05 --  2013 08 & 87 & \citet{Pilcher2014e} \\
      &                 & 2016-04-30.0        &  1 & Martin Lehk\'y \\
      &                 & 2016-04-30.9        &  1 & Martin Lehk\'y \\
      &                 & 2016-05-02.9        &  1 & Martin Lehk\'y \\
  608 &        Adolfine & 2006-10-14.9        &  1 & Raymond Poncy \\
      &                 & 2006-10-24.9        &  1 & Raymond Poncy \\
      &                 & 2006-10-26.9        &  1 & Raymond Poncy \\
      &                 & 2006-10-27.9        &  1 & Raymond Poncy \\
      &                 & 2013-02-03.9        &  1 & Roberto Crippa, Federico Manzini \\
      &                 & 2006-10-11.0        &  1 & Roberto Crippa, Federico Manzini \\
      &                 & 2006-10-15.9        &  1 & Roberto Crippa, Federico Manzini \\
      &                 & 2006-10-26.9        &  1 & Roberto Crippa, Federico Manzini \\
      &                 & 2016-08-31.9        &  1 & Martin Lehk\'y \\
      &                 & 2016-09-02.9        &  1 & Martin Lehk\'y \\
  616 &            Elly & 2010 01 --  2010 02 &  2 & \citet{Warner2010a} \\
      &                 & 2010 02 --  2010 02 &  2 & \citet{Durkee2010a} \\
      &                 & 2014 02 --  2014 02 &  2 & \citet{Parnafes2014} \\
      &                 & 2014 02 --  2014 02 &  4 & \citet{Stephens2014b} \\
      &                 & 2014 02 --  2014 03 &  8 & \citet{Klinglesmith2014} \\
      &                 & 2016-10-04.8        &  1 & Martin Lehk\'y \\
  775 &         Lumiere & 1983 05 --  1983 05 &  3 & \citet{Binzel1987a} \\
      &                 & 2001-09-20.0        &  1 & Ren\'e Roy \\
      &                 & 2001-09-26.0        &  1 & Ren\'e Roy \\
      &                 & 2001-09-31.0        &  1 & Ren\'e Roy \\
      &                 & 2001-08-27.1        &  1 & Ren\'e Roy \\
      &                 & 2003-03-23.9        &  1 & Claudine Rinner \\
      &                 & 2005-07-07.1        &  1 & Stephane Charbonnel, Pierre Dubreuil \\
      &                 &                     &    & Alain Lopez, Gilles Kober \\
      &                 & 2006-10-15.0        &  1 & Raymond Poncy \\
      &                 & 2006-10-24.9        &  1 & Raymond Poncy \\
      &                 & 2006-10-26.9        &  1 & Raymond Poncy \\
      &                 & 2006-10-27.8        &  1 & Raymond Poncy \\
      &                 & 2016-09-25.9        &  1 & Martin Lehk\'y \\
  822 &          Lalage & 1992 09 --  1992 09 &  2 & \citet{Wisniewski1997} \\
      &                 & 2009 10 --  2009 10 &  2 & \citet{Higgins2011a} \\
      &                 & 2014 01 --  2014 01 &  2 & \citet{Stephens2014b} \\
      &                 & 2014 02 --  2014 03 &  6 & \citet{Klinglesmith2014} \\
      &                 & 2016-12-29.0        &  1 & Martin Lehk\'y \\
  918 &            Itha & 2011 06 --  2011 06 &  4 & \citet{Oey2012b} \\
      &                 & 2011 06 --  2011 07 &  3 & \citet{Folberth2012} \\
      &                 & 2016-08-24.0        &  1 & Martin Lehk\'y \\
      &                 & 2016-12-03.8        &  1 & Martin Lehk\'y \\
  955 &         Alstede & 2016-03-17.9        &  1 & Martin Lehk\'y \\
 1095 &          Tulipa & 1983 02 --  1983 02 &  1 & \citet{Binzel1987a} \\
      &                 & 2005-04-15.0        &  1 & Ond\v rejov \\
      &                 & 2005-04-29.0        &  1 & Ond\v rejov \\
      &                 & 2005-04-28.0        &  1 & Pierre Antonini, Raoul Behrend \\
      &                 & 2005-05-15.0        &  1 & Pierre Antonini, Raoul Behrend \\
      &                 & 2005-05-26.9        &  1 & Pierre Antonini, Raoul Behrend \\
      &                 & 2005-05-32.0        &  1 & Pierre Antonini, Raoul Behrend \\
      &                 & 2005-06-02.0        &  1 & Pierre Antonini, Raoul Behrend \\
      &                 & 2005-06-01.9        &  1 & Gino Farroni \\
      &                 & 2005-05-32.0        &  1 & Gino Farroni \\
      &                 & 2016-06-24.0        &  1 & Martin Lehk\'y \\
 1219 &          Britta & 1983 09 --  1983 12 & 12 & \citet{Binzel1987a} \\
      &                 & 2014 01 --  2014 01 &  4 & \citet{Stephens2014b} \\
      &                 & 2014 02 --  2014 03 &  4 & \citet{Klinglesmith2014} \\
      &                 & 2016-10-30.9        &  1 & Martin Lehk\'y \\
 1251 &          Hedera & 2007 07 --  2007 10 &  9 & \citet{Oey2008b} \\
      &                 & 2016-08-08.0        &  1 & Martin Lehk\'y \\
 1320 &          Impala & 2006 05 --  2006 05 &  3 & \citet{Warner2006c} \\
      &                 & 2016 03 --  2016 04 &  5 & \citet{Benishek2016} \\
      &                 & 2016-05-08.0        &  1 & Martin Lehk\'y \\
 1380 &         Volodia & 2016-09-24.0        &  1 & Martin Lehk\'y \\
 1663 &     van den Bos & 2010 09 --  2010 11 & 32 & \citet{Stephens2011c} \\
      &                 & 2010 10 --  2010 11 &  4 & \citet{Ruthroff2011} \\
      &                 & 2016-05-12.0        &  1 & Martin Lehk\'y \\
 2511 &       Patterson & 2016-12-03.9        &  1 & Martin Lehk\'y \\
34817 &    Shiominemoto & 2003 07 --  2003 07 &  2 & \citet{Warner2004a} \\
      &                 & 2006 12 --  2006 12 &  2 & \citet{Warner2007c} \\
      &                 & 2011 11 --  2011 11 &  3 & \citet{Warner2012a} \\
      &                 & 2015 02 --  2015 02 &  4 & \citet{Stephens2015c} \\
      &                 & 2016-08-31.0        &  1 & Martin Lehk\'y \\ \hline
\end{longtable}
\twocolumn

\onecolumn
\begin{landscape}
\begin{longtable}{lll}
\caption{\label{tab:observatories}List of observers, their locations and telescope specifications.}\\
\hline
 Observer & Observatory & Telescope specification \\ \hline\hline

\endfirsthead
\caption{continued.}\\

\hline
 Observer & Observatory & Telescope specification \\ \hline\hline
\endhead
\hline
\endfoot
 Roberto Crippa      &  Tradate, Italy (B13)                                   & Reflector 0.65m, F/D=5, Apogee Alta 1001, KAF-1001ME \\
 Federico Manzini    &  Sozzago, Italy (A12)                                   & D=0.40m F/D=5, Hisis 33 \\
 Martin Lehk\'y      &  Ond\v rejov Observatory, Czech Republic (557)          & BlueEye 600 \\
 Ren\'e Roy          &  Observatoire de Blauvac, Blauvac, France  (627)        & D=0.312m, Hisis 22, KAF-400  \\
 Julian Oey          &  Kingsgrove, NSW, Australia (E19)                       & D=0.25m Schmidt-Cassegrain, F/D-5.2, SBIG ST-402ME CCD  \\
 Raymond Poncy       &  Le Cr\'es Observatory, Le Cr\'es, France (177)         & D=0.20m F/D=3.3  \\
 Claudine Rinner     &  224 Ottmarsheim, France (224)                          & D=0.305m F/D=10, ST8e, KAF1602E, KAF3200me  \\
 Stephane Charbonnel &  Observatoire de Durtal, Durtal, France (949)           & D=0.30m, F/D=3.5, KAF-400e \\
 Pierre Antonini     &  Observatoire des Hauts Patys, B\'edoin, France (132)   &  D=0.305m, F/D=2.6 , NJP160, MCMT2, KAF-400 \\
 \hline
\end{longtable}
\end{landscape}
\twocolumn

\bibliography{mybib,bibliography_all}
\bibliographystyle{model2-names}



\end{document}